# A Comprehensive Study of CRM through Data Mining Techniques

## Md. Rashid Farooqi[1] and Khalid Raza[2]

Department of Commerce & Business Studies[1], Department of Computer Science[2]
Jamia Millia Islamia (Central University), New Delhi, INDIA
prashidmgr@gmail.com [1], kraza@jmi.ac.in[2]

**ABSTRACT**
*In today's competitive scenario in corporate world, "Customer Retention" strategy in Customer Relationship Management (CRM) is an increasingly pressed issue. Data mining techniques play a vital role in better CRM. This paper attempts to bring a new perspective by focusing the issue of data mining applications, opportunities and challenges in CRM. It covers the topic such as customer retention, customer services, risk assessment, fraud detection and some of the data mining tools which are widely used in CRM.*

## 1. INTRODUCTION

### 1.1 Customer Relationship Management (CRM)
The emergence of Information Technology and use of computer in every field of activities has created a new buzz in the field of marketing and that is the concept of Consumer Relationship Management (CRM). The concept of CRM defined as *"the process of acquiring, retaining and growing profitable customer which requires a clear focus on service attributes that represent value to the customer and creates loyalty"*. The CRM is a term applied to processes implemented by company to handle their contact with their customers. CRM software is used to support these processes, storing information about prospective customers.

The term CRM is generally used to refer to software based approach for handling customer relationship. Most CRM software vendors stress that a successful CRM strategy require a holistic approach. CRM initiative often fails because implementation was limited to software installation without providing the appropriate motivation for employees to learn, provide inputs and take full advantage of the information system. The customer relationship is neither a concept nor a project, instead a business strategy that aims to understand, anticipate and manage the needs of an organization's current and potential customer.

CRM includes many-aspects which relate directly to one another:

- Front office operations: Direct interaction with customer, e.g. face to face, email, online services, phone calls etc.
- Back office operations: Operations that ultimately affect the activities of the front office e.g. billing, maintenance, marketing, planning, finance, manufacturing, advertising, etc.
- Business Relationship: Interaction with other companies and partners, such as suppliers/vendors and retail outlets, distributors, industry networks. This external network supports front and back office activities.
- Analysis: Key CRM data can be analyzed in order to plan target-marking campaigns, conceive business strategies, and judge the success of CRM activities e.g. market share, number and type of customers, revenue, profitability, etc. [1]

### 1.2 Technological Consideration
The basic building blocks of CRM are:

*Customer Databases:* A database for customer life cycle information about each customer and prospect and their interactions with the organization, including order information, support information, requests, complaints, interviews and survey responses.

*Customer Intelligence:* Translation customer needs and profitability projection into game plans for different segments or groups of customers, captured by customer interactions into software that tracks whether that game plan is followed and whether the desired outcomes are obtained.

*Business Modeling Customer Relationship Strategy, Goals and Outcomes:* Numbers and description of whether goals were met and models of customer segments and game plans worked as hypothesized.

*Learning and Competency Management Systems:* Training and improving processes and technology that enable the organizations to get closer to achieving the desired results. Complex system require practice in order to achieve desired outcomes, especially when human and technology are interacting. Iteration is the key to refining, improving and innovating to stay ahead of the competition in CRM.

### 1.3 Customer Life Cycle (CLC)
The customer life cycle has three stages:

(i) *Customer acquisition:* Customer acquisition is the major objective of all organization. Different organization provides services in order to increase their customer databases. Customer acquisition strategy focuses upon making the buyer perspective customer in future. The sources of acquisition may be through enquiries, suspects, former customers, existing customers, competitors, etc.

(ii) *Retaining good customers:* It is the process of keeping customer in the customer inventory for an unending period by meeting the needs and exceeding the expectations of those customers. Retaining customer is more important than attracting new customer. Customer retention is all about customer loyalty. It enables a long term relationship of mutual benefits both to the organization and customer concerns.

(iii) *Making the relationship of customers:* It is the process of acquiring, retaining, maintaining and growing profitable customers. The aim of marketing is to meet and satisfy target customer needs and wants. The modern marketing concept makes customer the centre stage of organization effort the focus within their marketing concepts to reach the target customers. So customer is considered to be the king forever in all markets.

**1.4 Scope of CRM**

The scope of customer relationship management in terms of point is as follows:

- Implementing appropriate systems to support customer knowledge acquisition, sharing and measuring CRM effectiveness.

- Integrating the activities of marketing, sales and service to achieve a common goal.

- Applying customer knowledge to continuously improve performance through a process of learning from successes and failures.

- Acquiring and continuously updating knowledge about customer needs, motivations and behavior over the lifetime of the relationship.

- Measuring both inputs across all functions including marketing, sales and service costs and outputs in terms of customer revenue, profit and value.

- Constantly flexing the balance between marketing, sales and service inputs against changing customer needs to maximize profit.

**1.5 Benefits of CRM**

The benefits of CRM are:

- CRM permits business to leverage information from their databases to achieve customer retention and to cross sell new products.

- Companies that implement CRM make better relationships with their customers achieve loyal customers and a substantial payback increased revenues and reduced cost.

- Low maintenance and expansion cost owing to the use of modern administration tool which allow bank employee to make a wide range of modification to the system.

- CRM focus upon profitable client through efficient segmentation according to individual behavior.

- CRM results both in higher revenue and lower cost making companies more effective and efficient.

**2. DATA MINING**

Data mining has attracted a great attention in the information industry and in society as a whole in recent years. Due to the availability of huge amounts of data and the imminent need for turning such data into useful information and knowledge. The information and knowledge gained can be used for applications ranging from market analysis, fraud detection and customer retention etc. [2]

Data mining (DM) refers to extracting or "mining" knowledge from large amounts of data [11]. DM is the science of finding new interesting patterns and relationship in huge amount of data. Data Mining is defined as "the process of discovering meaningful new correlations, patterns, and trends by digging into large amounts of data stored in warehouses". Data mining is not specific to any industry. It requires intelligent technologies and the willingness to explore the possibility of hidden knowledge that resides in the data. [4]

**2.1 Data Mining Application Domains**

From the CRM point of view, the data mining applications include but not limited to the following:

*Customer Retention*: Sophisticated customer-retention programs begin with modeling those customers who have defected to identify patterns that led to their defection. These models are then applied to the current customers to identify likely defectors so that preventive actions can be initiated.

*Sales and Customer Services:* In today's highly competitive environment, superior customer service creates the sales leaders. When information is properly aggregated and delivered to front-line sales and service professionals, customer service is greatly enhanced. If customer information is available, rule-based software can be employed to automatically recommend products. The programs like market-basket analysis have already shown phenomenal gains in cross-selling ratios, floor and shelf layout and product placement improvements and better layout of catalog and web pages.

*Marketing:* Marketing depends heavily on accurate information to execute retention campaigns, lifetime value analysis, trending targeted promotions, etc. Only by having a complete customer profile can promotions be targeted and targeting dramatically increase response rates and thus decreases campaign cost.

*Risk Assessment & Fraud Detection:* An accessible customer base significantly reduces the risk of entering into undo risk. For example, a bank can identify fiscally related companies that may be in financial jeopardy before extending a loan to them.

**2.2 Data Mining Techniques for CRM**

Data mining techniques deal with discovery and learning. Data mining techniques may be helpful to accomplish the goal of CRM by extracting or detecting hidden customer characteristics and behaviours from large databases. Following are the some of the popular data mining techniques:

## Association Rule Learning

Association Rule Learning is a popular method for discovering interesting relations between variables in large databases. Agrawal *et al.* [1,5] introduced association rules for discovering regularities between products in large scale transaction data recorded by point-of-sale (POS) systems in supermarkets. For example,

$$\{Computer, Minotor\} \Rightarrow \{U.P.S.\}$$

The above association rule found in the sales data of a supermarket would indicate that if a customer buys computers and monitors together, he or she is likely to also buy a U.P.S. This type of information can be used for decision making about marketing activities such as, promotional pricing or product placements, market basket analysis, etc.

## Classification & Prediction

Classification and prediction are two forms of data analysis that can be used to extract models describing important data classes or to predict future data trends. It aims at building a model to predict future customer behaviors through classifying databases records into a number of predefined classes based on certain criteria. Classification predicts categorical (unordered) labels, prediction models continuous valued functions [2].

Basic techniques for data classification are decision tree classifier, Bayesian classifier, Bayesian belief networks, rule-based classifiers, and support vector machines. Methods for prediction include linear regression, non-linear regression, etc.

## Clustering

Clustering is the method by which similar type of records are grouped together. Usually, clustering is done to give the end user a high-level view of what is going on in the database. Clustering is useful for coming up with a birds-eye view of the business [4].

## Regression Analysis

Regression analysis helps us understand how the typical value of the dependent variable changes when any one of the independent variables is varied, while the other independent variables are held fixed. Regression analysis is widely used for prediction and forecasting.

## Visualization

Visualization refers to the presentation so that users can view complex patterns. According to Friedman (2008) the main goal of data visualization is to communicate information clearly and effectively through graphical means. It is used with other data mining models to provide a better and clearer understanding of the discovered patterns or relationships.

## 2.3 Taxonomy of Data Mining Tools

Today there are various data mining tools available in the market. These tools can be broadly placed in following three categories (i) general purpose tools (ii) integrated DSS / OLAP / DM tools and (iii) application specific tools (see Table 1).

Table 1: Categorization of available DM tools

| GENERAL PURPOSE TOOLS |
| --- |
| SAS Enterprise Miner<br>www.sas.com/technologies/analytics/datamining/miner |
| IBM Intelligent Miner<br>www.01.ibm.com/software/data/iminer |
| Unica Pattern Recognition Workbench<br>http://www.unica.com |
| IBM SPSS Modeler<br>www.spss.com/software/modeling/modeler-pro |
| Ghost Miner<br>http://cncmining.com/ |
| XLMiner<br>www.resample.com/xlminer/download.shtml |
| CART & MARS<br>www.salford-systems.com/ |
| SGI Mineset<br>www.sgi.com/ |
| Oracle Darwin<br>www.oracle.com/technology/documentation/darwin.html |
| Angoss Knowledge Seeker<br>http://www.angoss.com/ |
| Weka<br>www.cs.waikato.ac.nz/ml/weka |
| Rapid Miner<br>http://rapid-i.com/ |
| TIBCO Spotfire S+<br>http://spotfire.tibco.com/ |
| **INTEGRATED DSS/OLAP/DM TOOLS** |
| Cognos Scenario<br>www.cognos.com/busintell/products/index.html |
| Business Objects<br>www.sap.com/solutions/sapbusinessobjects/index.epx |
| **APPLICATION-SPECIFIC TOOLS** |
| KD1 (Knowledge Discovery One)<br>www.kd1.com |
| ESTARD Data Miner<br>http://www.estard.com/ |
| Unica Detect<br>http://www.unica.com/ |
| Unica Leaders<br>http://www.unica.com/ |
| Unica Predictive Insight<br>http://www.unica.com/ |

The general purpose tools contain a large segment of the market and are non-application specific. Some of the examples are, SAS Enterprise Miner, IBM Intelligent Miner, Unica Pattern Recognition Workbench (PRW), IBM SPSS Modeler, Ghost Miner, XLMiner, CART & MARS, SGI Mineset, Oracle Darwin, Angoss Knowledge Seeker, Weka, Rapid Miner etc.

The integrated DSS/OLAP/DM tool addresses a very real and compelling business requirement of having a single multi-functional decision-support tool that can provide management reporting, online analytical processing and data mining capabilities within a common framework. Some of the examples in this category are: Cognos Scenario, Business Objects, etc.

The application-specific tools are rapidly gaining momentum. It offers business solution rather than a technology searching for solution. Some of the tools under this category are: KD1 (focuses on retail), ESTARD Data Miner (focuses on the insurance industry & marketing), Unica Detect, Unica Leaders, Unica Predictive Insight (focuses on fraud detection, marketing) etc.

## 3. DATA MINING CHALLENGES & OPPORTUNITIES IN CRM

Following are the key data mining challenges and opportunities for better customer relationship management:

- *Non-trivial results almost always need a combination of DM techniques:* Chaining/composition of DM, and more generally data analysis, operations is important. In order to analyze CRM data, one needs to explore the data from different angles and look at its different aspects. This should require application of different types of DM techniques and their application to different "slices" of data in an interactive and iterative fashion. Hence, the need to use various DM operators and combine (chain) them into a single "exploration plan".

- *There is a strong requirement for data integration before data mining:* In both cases, data comes from multiple sources. For example in CRM, data needed may come from different departments of an organization. Since many interesting patterns span multiple data sources, there is a need to integrate these data before an actual data mining exploration can start.

- *Diverse data types are often encountered, which requires the integrated mining of diverse and heterogeneous data:* In CRM, while dealing with this issue is not critical, it is nonetheless important. Customer data comes in the form of structured records of different data types (e.g., demographic data), temporal data (e.g., weblogs), text (e.g., emails, consumer reviews, blogs and chat-room data), (sometimes) audio (e.g., recorded phone conversations of service reps with customers).

- *Highly and unavoidably noisy data must be dealt with:* In CRM, weblog data has a lot of "noise" (due to crawlers, missed hits because of the caching problem, etc.). Other data pertaining to customer "touch-points" has the usual cleaning problems seen in any business-related data.

- *Real-world validation of results is essential for acceptance:* In CRM, as in many DM applications, discovered patterns are often treated as hypotheses that need to be tested on new data using rigorous statistical tests for the actual acceptance of the results. This is even more so for taking or recommending actions, especially in such high-risk applications as in the financial and medical domains. Example: recommending investments to customers (it is actually illegal in the US to let software give investment advice).

- *Developing deeper models of customer behavior:* One of the key issues in CRM is how to understand customers.

Current models of customers mainly built based on their purchase patterns and click patterns at web sites. Such models are very shallow and do not have a deep understanding of customers and their individual circumstances. Thus, many predictions and actions about customers are wrong. It is suggested that information from all customer touch-points be considered in building customer models. Marketing and psychology researchers should also be involved in this effort. Two specific issues need to be considered here. First, what level should the customer model be built at, namely at the aggregate level, the segment level, or at the individual level? The deciding factor is how personalized the CRM effort needs to be. Second is the issue of the dimensions to be considered in the customer profile. These include demographic, psychographic, macro-behavior (buying, etc.), and micro-behavior (detailed actions in a store, e.g. individual clicks in an online store) features.

- *Acquiring data for deeper understanding in a non-intrusive, low-cost, high accuracy manner:* In many industrial settings, collecting data for CRM is still a problem. Some methods are intrusive and costly. Datasets collected are very noisy and in different formats and reside in different departments of an organization. Solving these pre-requisite problems is essential for data mining applications.

- *Managing the "cold start/bootstrap" problem:* At the beginning of the customer life cycle little is known, but the list of customers and the amount of information known for each customer increases over time. In most cases, a minimum amount of information is required for achieving acceptable results (for instance, product recommendations computed through collaborative filtering require a purchasing history of the customer). Being able to deal with cases where less than this required minimum is known is a therefore a major challenge.

## 4. CONCLUSION

Data mining is a growing discipline which originated outside statistics in the database management community, mainly for commercial concerns. Data mining can be considered as the branch of exploratory statistics where one tries to find new and useful patterns, through the extensive use of classic and new algorithms.

Application of customer relationship management tool in business gives a new dimension. It proved beneficial but applying data mining in customer relationship management was further more beneficial. Although the data mining tools market is relatively small, at the same time the data mining application solution market is growing exponentially. Our main focus was on customer retention techniques to enhance our customer relationships via Data Mining. Data Mining would fasten up the process of searching large databases so as to extract customer buying patterns, to classify customers into groups which also make databases to be handled efficiently.

# 6. REFERENCES

[1] www.wikipedia.org

[2] Han and Kamber, *"Data Mining concepts and techniques"*, Morgan Kaufmann Publishers, 2006

[3] Herb Edelstein, *"Building Profitable Customer Relationship with Data Mining"*, President, Two Crows Corporation.

[4] Alex Berson, Stephen Smith & Kurt Threaling, *"Building Data Mining Application for CRM"*, Tata McGraw Hill.

[5] R. Agrawal; T. Imielinski; A. Swami, "Mining Association Rules Between Sets of Items in Large Databases", *SIGMOD Conference 1993*, pp. 207-216.

[6] V. Kumar and Werner Reinartz, *"Customer Relationship Management: A Databased Approach"*, John Wiley & Sons, Inc.

[7] Michael J.A. Berry and Gordon Linoff, *"Data Mining Techniques: For Marketing, Sales and Customer Support"*, John Wiley & Sons, Inc..

[8] V. Dhar, R. Stein, *"Seven Methods for Transforming Corporate Data into Business Intelligence"*, Prentice Hall of India.

[9] C. Westphal, T. Blazton, Chris Westphal, *"Data Mining Solutions: Methods and Tools for Solving Real-World Problems"*, John Wiley & Sons, Inc.

[10] D. J. Hand, H. Mannila and P. Smyth, *"Principles of Data Mining"*, MIT Press.

[11] Khalid Raza, "Application of Data Mining in Bioinformatics", *Indian Journal of Computer Science and Engineering*, Vol.1, No. 2, pp. 114-118, 2010.